\documentclass[aps,prc,twocolumn,superscriptaddress,floatfix, nofootinbib,preprintnumbers]{revtex4-1}
\usepackage{graphicx}  
\usepackage{dcolumn}   
\usepackage{bm,relsize}        
\usepackage{amssymb, amsmath}
\usepackage{textcomp}
\usepackage{wasysym}
\usepackage{slashed}
\usepackage{braket}
\usepackage{ulem}
\usepackage{caption, subcaption}
\usepackage{multirow}

\usepackage{lipsum, color}
\usepackage[usenames,dvipsnames,svgnames]{xcolor}
\usepackage{hyperref}
\hypersetup{colorlinks, citecolor=blue, linkcolor=blue, urlcolor=blue}
\usepackage{booktabs}
\usepackage[utf8]{inputenc}

\def\beq{\begin{equation}}
\def\eeq{\end{equation}}
 \def\be{\begin{equation}} \def\ee{\end{equation}}
\def\bea{\begin{eqnarray}} \def\eea{\end{eqnarray}}

\usepackage{caption}
\usepackage{subcaption}
\usepackage{multirow}
\allowdisplaybreaks

\begin{document}

\title{Distinguishing charged lepton flavor violation scenarios with inelastic $\mu\rightarrow e$ conversion}

\author{W. C. Haxton}
\affiliation{Department of Physics, University of California, Berkeley, CA 94720, USA}
\affiliation{Lawrence Berkeley National Laboratory, Berkeley, CA 94720, USA}
\author{Evan Rule}
\affiliation{Department of Physics, University of California, Berkeley, CA 94720, USA}
\affiliation{Theoretical Division, Los Alamos National Laboratory, Los Alamos, NM 87545, USA}

\date{\today}

\preprint{N3AS-24-014, LA-UR-24-24058}

\begin{abstract}
The Mu2e and COMET experiments are expected to improve existing limits on charged lepton flavor violation (CLFV) by roughly four orders of magnitude. $\mu\rightarrow e$ conversion experiments are typically optimized for electrons produced without nuclear excitation, as this maximizes the electron energy and minimizes backgrounds from the free decay of the muon. Here we argue that Mu2e and COMET will be able to extract additional constraints on CLFV from inelastic $\mu \rightarrow e$ conversion, given the $^{27}$Al target they have chosen and backgrounds they anticipate.  We describe CLFV scenarios in which inelastic CLFV can induce measurable distortions in the near-endpoint spectrum of conversion electrons, including cases where certain contributing operators cannot be probed in elastic $\mu \rightarrow e$ conversion.  We extend the nonrelativistic EFT treatment of elastic $\mu \rightarrow e$ conversion to include the new nuclear operators needed for the inelastic process, evaluate the associated nuclear response functions, and describe several new-physics scenarios where the inelastic process can provide additional information on CLFV.

\end{abstract}


\maketitle
The observation of neutrino oscillations implies that flavor violation also occurs among the charged leptons, though the neutrino-mediated contribution to charged lepton flavor violation (CLFV) is unobservable, suppressed by the small neutrino mass.    
Consequently, observation of CLFV would be evidence of additional new physics \cite{BARBIERI1994212,Bernstein:2013hba,Calibbi:2017uvl}.  One of the most sensitive tests of CLFV is $\mu \rightarrow e$ conversion, in which muons are stopped in a target, capture into
the 1s Coulomb orbits of the target nuclei, and convert into mono-energetic electrons,
unaccompanied by any other final-state leptons.
%

The quantity reported by $\mu\rightarrow e$ conversion experiments is the branching ratio 
\begin{equation}
    R_{\mu e}=\frac{\Gamma\left[\mu^-+(A,Z)\rightarrow e^-+(A,Z)\right]}{\Gamma\left[\mu^-+(A,Z)\rightarrow \nu_\mu + (A,Z-1)\right]},
\end{equation}
where the numerator is the rate for the CLFV process and the denominator is the rate for standard muon capture. Currently, the best limit is $R_{\mu e}< 7.0\times 10^{-13}$ at 90\% confidence level, obtained from SINDRUM II measurements using a gold target \cite{SINDRUMII:2006dvw}. Two new experiments now in construction, Mu2e \cite{Bernstein_2019} at Fermilab and COMET 
\cite{10.3389/fphy.2018.00133} at J-PARC, are expected to improve existing limits by about four orders of magnitude, reaching a single-event sensitivity $R_{\mu e}\lesssim 10^{-17}$ for a $^{27}$Al target.

The momentum of the conversion electron (CE) produced from a muon in a $1s$ atomic orbit is given by
\begin{equation}
\vec{q}^{\,2}=\frac{M_T}{m_\mu + M_T}\left[\left(m_{\mu}-E^\mathrm{bind}_{\mu}-\Delta E_\mathrm{nuc}\right)^2-m_e^2\right], 
\label{eq:inelastic_energy}
\end{equation}
where $M_T$ is the mass of the nuclear target, $\Delta E_\mathrm{nuc}=E_f-E_i$ is the energy gap between the final and initial nuclear states, and $E_\mu^\mathrm{bind}\approx 0.463$ MeV is the muon binding energy for $^{27}$Al. The CLFV signal is a mono-energetic, relativistic electron with energy $E_\mathrm{CE}\approx m_\mu$, which must be distinguished from background electrons originating from the standard-model decay in orbit $\mu\rightarrow e+2\nu$ (DIO). Near the endpoint, the spectrum of DIO electrons is suppressed by $(E-E_\mathrm{end})^5$, where $E_\mathrm{end}$ is given by $E_\mathrm{CE}$ when $\Delta E_\mathrm{nuc}=0$. In order to minimize the DIO background, experiments typically focus on the most energetic CEs, those produced in \textit{elastic} conversion where $\Delta E_\mathrm{nuc}=0$.

If a nonzero elastic rate is observed in an experiment, it will establish the existence of CLFV, but provide no information about the underlying source: The coefficient of any candidate operator can be dialed to reproduce the signal. To learn more, one would have to perform additional experiments that employ target nuclei with varying ground-state spin, isospin, valence nucleon, and spin-orbit structure \cite{Haxton:2022piv} --- a tedious process.  Here we argue that a single experiment can yield multiple independent constraints on CLFV, provided the target has specific properties.  The target chosen by Mu2e and COMET, $^{27}$Al, is an outstanding example of such a target.


Specifically, we consider the case of \textit{inelastic} $\mu\rightarrow e$ conversion, where the nucleus transitions to an excited final state\footnote{Previous studies \cite{PhysRevLett.3.111,KOSMAS1990641,CHIANG1993526,KOSMAS1994637,PhysRevC.62.035502,KOSMAS2001443,PhysRevC.99.065504}, which frequently refer to this scenario as \textit{incoherent} $\mu\rightarrow e$ conversion, have estimated the total response to all excited states (for selected CLFV operators), in contrast to our present focus on experimentally accessible low-lying transitions and CLFV operators for which inelastic transitions can provide unique information.}. Provided one focuses on transitions where $\Delta E_\mathrm{nuc} \lesssim 2$ MeV, signals can appear above the DIO background, distinguishable from the elastic response, given the anticipated momentum resolution of Mu2e and COMET. This inelastic contribution probes certain CLFV operators that cannot contribute to elastic $\mu \rightarrow e$ conversion,  due to the parity (P) and time-reversal (T) selection rules.  Further, because the inelastic and elastic contributions can be separated by doing a shape analysis in the endpoint region where the DIO background is low, the inelastic contribution provides additional information, regardless of the operator origin of CLFV.


$^{27}$Al is an interesting case for inelastic $\mu \rightarrow e$ conversion. As a relatively light nucleus with an unpaired nucleon --- the na\"ive description of the $J^\pi=5/2^+$ ground state is a proton hole in a closed $1d_{5/2}$ shell --- $^{27}$Al's low-energy spectrum 
includes three reasonably well spaced excited states in the endpoint region, $J^{\pi}=1/2^+$ (0.844 MeV), $3/2^+$ (1.015 MeV), and $7/2^+$ (2.212 MeV).  Transitions to the second and third states are ``allowed" (requiring angular momentum transfer of $|J| \le$ 1).  All three states can be excited by operators whose
strengths are consistent with current bounds on CLFV, producing signals that can be distinguished from background.  The contribution from DIO is modest and, as this background 
is well understood \cite{PhysRevD.84.013006,SZAFRON201661,PhysRevD.94.051301}, can be subtracted in a shape analysis.  The less well understood electron background from radiative muon capture (RMC) does not contribute in this energy window, as the endpoint for RMC electrons in $^{27}$Al is 3.62 MeV below the maximum CE energy.



\textit{Experimental signal.} A mono-energetic CE will be registered in the Mu2e detector with a reconstructed momentum $q_\mathrm{rec}$ that differs from the initial momentum $q$ due to energy losses as the electron moves through additional layers of the aluminium target and proton absorber, before reaching the calorimeter \cite{Mu2e:2014fns}. Such target effects dominate over those due to the intrinsic resolution of the Mu2e detector. Recently, the Mu2e collaboration performed detailed Monte Carlo (MC) simulations in order to characterize the detector response and overall efficiency \cite{Mu2e:2022ggl}. Based on this study, the energy loss experienced by electrons in the signal window ($q\gtrsim 100$ MeV) results in a typical shift in the reconstructed momentum $\delta q_0 = q_\mathrm{rec}-q= -0.5497(26)$ MeV/c. The smearing is highly asymmetric, with a long low-energy tail. 



The shape of the $\mu\rightarrow e$ conversion signal generated in a single nuclear transition does not depend on the underlying CLFV mechanism: a mono-energetic electron is produced with momentum $q$, given by Eq. \eqref{eq:inelastic_energy}, with this signal then smeared out by the energy losses described above. But when multiple nuclear final states contribute, the combined signal can be quite sensitive to the underlying source of the CLFV, reflecting operator-dependent relative branching ratios.  This is the source of the additional information not available
from the elastic transition alone.

In Refs. \cite{Rule:2021oxe,Haxton:2022piv}, the most general expression for the elastic $\mu\rightarrow e$ conversion rate was derived in non-relativistic effective theory (NRET), yielding the factorized form 
\begin{equation}
    \Gamma_{\mu e}(gs\rightarrow gs)=\Gamma_0 \sum_{\tau=0,1}\sum_{\tau'=0,1}\sum_i \tilde{R}_{i}^{\tau\tau'}W^{\tau\tau'}_i,
    \label{eq:elast}
\end{equation}
where the CLFV response functions $\tilde{R}_i^{\tau\tau'}$ are dimensionless, target- and transition-independent coefficients that encode all available information about the underlying CLFV mechanism.  They are bilinear functions of the coefficients of the 16 CLFV nucleon-level NRET operators defined in Table \ref{tab:response_properties}. The $W_i^{\tau\tau'}$ are dimensionless nuclear response functions that depend on both target and transition. Subscripts $i$ and superscripts $\tau,\tau'$ indicate the
operator and isospin dependence, respectively.  Details can be found in \cite{Haxton:2022piv}, where the NRET is developed through
linear order in bound nucleon $\vec{v}_N$ and muon $\vec{v}_\mu$ velocities.
 
The NRET construction of the single-nucleon CLFV transition operator is fully general, applicable to both elastic and inelastic transitions. It generates a hierarchy of low-energy operators, organized according to available dimensionless small parameters $q R > |\vec{v}_N| > |\vec{v}_\mu|$, where $R$ is the nuclear size. Experimental results constrain the coefficients of the NRET operators.  Once obtained, these constraints can be ported to higher energies, through a process of matching successive effective theories \cite{Haxton:2024lyc}. Which nuclear responses arise and how they are related to NRET operators depends on whether the $\mu \rightarrow e$ conversion is elastic or inelastic: Nuclear selection rules associated with parity and time-reversal invariance restrict the elastic response functions appearing in Eq. (\ref{eq:elast}), limiting the number of constraints $\tilde{R}_i^{\tau \tau^\prime}$ that can be imposed on underlying CLFV operator coefficients.

\begin{table}[]
    \centering
       {\renewcommand{\arraystretch}{1.3}
    \begin{tabular}{lccccl}
     \hline
      \hline
      $\mathcal{O}_1=1_L1_N$ & \multicolumn{2}{c}{$\mathcal{O}_2'=i\hat{q}\cdot\vec{v}_N$}  & \multicolumn{3}{c}{$\mathcal{O}_3=i\hat{q}\cdot[\vec{v}_N\times\vec{\sigma}_N]$} \\
      $\mathcal{O}_4=\vec{\sigma}_L\cdot\vec{\sigma}_N$ & \multicolumn{3}{c}{$\mathcal{O}_5=\vec{\sigma}_L\cdot(i\hat{q}\times\vec{v}_N)$} & \multicolumn{2}{c}{$\mathcal{O}_6=i\hat{q}\cdot\vec{\sigma}_L\;i\hat{q}\cdot\vec{\sigma}_N$}\\
      $\mathcal{O}_7=\vec{v}_N\cdot\vec{\sigma}_N$ & \multicolumn{2}{c}{$\mathcal{O}_8=\vec{\sigma}_L\cdot\vec{v}_N$} & \multicolumn{3}{c}{$\mathcal{O}_9=\vec{\sigma}_L\cdot(i\hat{q}\times\vec{\sigma}_N$)} \\
      $\mathcal{O}_{10}=i\hat{q}\cdot\vec{\sigma}_N$ & \multicolumn{2}{c}{$\mathcal{O}_{11}=i\hat{q}\cdot\vec{\sigma}_L$}  & \multicolumn{3}{c}{$\mathcal{O}_{12}=\vec{\sigma}_L\cdot(\vec{v}_N\times\vec{\sigma}_N$)} \\
      \multicolumn{3}{l}{$\mathcal{O}_{13}'=\vec{\sigma}_L\cdot(i\hat{q}\times[\vec{v}_N\times\vec{\sigma}_N])$} & \multicolumn{3}{c}{$\mathcal{O}_{14}=i\hat{q}\cdot\vec{\sigma}_L\;\vec{v}_N\cdot\vec{\sigma}_N$} \\
      \multicolumn{3}{l}{$\mathcal{O}_{15}=i\hat{q}\cdot\vec{\sigma}_L\;i\hat{q}\cdot[\vec{v}_N\times\vec{\sigma}_N]$} & \multicolumn{3}{c}{$\mathcal{O}_{16}'=i\hat{q}\cdot\vec{\sigma}_L\;i\hat{q}\cdot\vec{v}_N$} \\
      \hline
    Projection & Resp. & Range & Even J & Odd J & NRET \\
    \hline
       $\mathcal{M}_{JM}\left(1_N\right)$ & $M_{JM}$  & $J\geq 0$ & \textbf{E-E} & O-O & $\mathcal{O}_1,\mathcal{O}_{11}$\\
       $\mathcal{M}_{JM}\left(\vec{v}_N\cdot\vec{\sigma}_N\right)$ & $\tilde{\Omega}_{JM}$  & $J\geq 0$ &  O-E & E-O & $\mathcal{O}_7,\mathcal{O}_{14}$ \\
       $\mathcal{L}_{JM}\left(\vec{\sigma}_N\right)$ & $\Sigma^{\prime\prime}_{JM}$ & $J\geq 0$ & O-O & \textbf{E-E} & $\mathcal{O}_4,\mathcal{O}_6,\mathcal{O}_{10}$ \\
       $\mathcal{T}^\mathrm{mag}_{JM}\left(\vec{\sigma}_N\right)$ & $\Sigma_{JM}$ & $J\geq 1$ & E-O & O-E & $\mathcal{O}_4,\mathcal{O}_9$\\
       $\mathcal{T}^\mathrm{el}_{JM}\left(\vec{\sigma}_N\right)$ & $\Sigma^{\prime}_{JM}$ & $J\geq 1$ & O-O & \textbf{E-E} & $\mathcal{O}_4,\mathcal{O}_9$ \\
       $\mathcal{L}_{JM}\left(\vec{v}_N\right)$ & $\tilde{\Delta}^{\prime\prime}_{JM}$ & $J\geq 0$ & E-O & O-E & $\mathcal{O}_2,\mathcal{O}_8,\mathcal{O}_{16}$\\
       $\mathcal{T}^\mathrm{mag}_{JM}\left(\vec{v}_N\right)$  & $\Delta_{JM}$ & $J\geq 1$ & O-O & \textbf{E-E} & $\mathcal{O}_5,\mathcal{O}_8$\\
       $\mathcal{T}^\mathrm{el}_{JM}\left(\vec{v}_N\right)$  & $\Delta^{\prime}_{JM}$ & $J\geq 1$ & E-O & O-E & $\mathcal{O}_5,\mathcal{O}_8$\\
       $\mathcal{L}_{JM}\left(\vec{v}_N\times\vec{\sigma}_N\right)$ & $\Phi^{\prime\prime}_{JM}$ & $J\geq 0$ & \textbf{E-E} & O-O & $\mathcal{O}_3,\mathcal{O}_{12},\mathcal{O}_{15}$\\
       $\mathcal{T}^\mathrm{mag}_{JM}\left(\vec{v}_N\times\vec{\sigma}_N\right)$ & $\tilde{\Phi}_{JM}$ & $J\geq 1$ & O-E & E-O & $\mathcal{O}_{12},\mathcal{O}_{13}$\\
       $\mathcal{T}^\mathrm{el}_{JM}\left(\vec{v}_N\times\vec{\sigma}_N\right)$  & $\tilde{\Phi}^{\prime}_{JM}$ & $J\geq 1$ & \textbf{E-E} & O-O & $\mathcal{O}_{12},\mathcal{O}_{13}$\\
        \hline
    \hline
    \end{tabular}}
    \caption{Top: The 16 nucleon-level NRET operators formed from the electron velocity $\hat{q}$, the nucleon velocity $\vec{v}_N$, and the lepton/nucleon spin operators $\vec{\sigma}_L$ and $\vec{\sigma}_N$. Bottom: Multipole responses contributing to $\mu\rightarrow e$ conversion, their underlying charges/currents (in parentheses), their transformation properties under P-T (even E or odd O), and their associated NRET operators $\mathcal{O}_i$. The charge and longitudinal, transverse magnetic, and transverse electric current multipoles are denoted $\mathcal{M}$, $\mathcal{L}$, $\mathcal{T}^\mathrm{mag}$, and $\mathcal{T}^\mathrm{el}$, respectively. The six responses denoted E-E contribute to elastic $\mu \rightarrow e$ conversion, the others only to inelastic.}
    \label{tab:response_properties}
\end{table}

We have developed for the first time the NRET inelastic equivalent of Eq. (\ref{eq:elast}) and evaluated the associated nuclear form factors. This technical work, described in \cite{Haxton:2024amf}, extends the number of nuclear operators
from six to 11 (see Table \ref{tab:response_properties}).  Two new response functions --- associated with additional observables $\tilde{R}_i^{\tau \tau^\prime}$ --- are obtained, and three of the existing response functions acquire new terms, reflecting the 
absence of time-reversal invariance as a constraint on inelastic transitions.  

The two new responses are associated with the axial charge operator $\vec{v}_N \cdot \vec{\sigma}_N$ and the longitudinal projection of the convection current $\vec{q} \cdot \vec{v}_N$.  The former vanishes for elastic transitions because of P and T selection rules, the latter because of current conservation.

The three modified response functions consist of paired terms,
$(W_{\Sigma},W_{\Sigma'})$, $(W_{\Delta},W_{\Delta'})$, and $(W_{\tilde{\Phi}},W_{\tilde{\Phi}'})$, that arise from the transverse magnetic and electric 
projections of the same current and consequently probe the same CLFV interactions; that is,
$\tilde{R}_O^{\tau\tau'}=\tilde{R}_{O'}^{\tau\tau'}$ for $O=\Sigma,\Delta,\tilde{\Phi}$. In each pair,
one response is generated by a normal parity operator, $\pi=(-1)^J$ where $J$ is the multipolarity, the other by an abnormal
parity one, $(-1)^{J+1}$.  This can produce a lot of variability in the relative rates of elastic and inelastic $\mu \rightarrow e$ capture, as transitions to states of different $J$ (and potentially different parity) will sample different sets of multipole operators. Such variations can be  exploited as diagnostic tools, if both elastic and inelastic data are available.


We perform sensitivity studies of elastic and inelastic $\mu \rightarrow e$ conversion, exploring ``directions" in
parameter space defined by selected response functions.  We consider
\begin{equation}
    \frac{R_{\mu e}(gs\rightarrow f)}{R_{\mu e}(gs\rightarrow gs)}=\frac{\Gamma_{\mu e}(gs\rightarrow f)}{\Gamma_{\mu e}(gs\rightarrow gs)} \rightarrow \frac{W^{\tau\tau'}_{O}(gs\rightarrow f)}{W_{O}^{\tau\tau'}(gs\rightarrow gs)},
    \label{eq:rel_gamma_ratio}
\end{equation}
where on the right we take the limit of a single response function.  In this limit the CLFV particle
physics, which is independent of the transition, factors from the nuclear physics and drops out of the ratio in Eq. \eqref{eq:rel_gamma_ratio}.  The quantity we explore is nuclear, a response function ratio.  However, if the spectral endpoint signature of the chosen $W^{\tau \tau^\prime}_O$ is distinctive, the presence or absence of this signature
would constrain $\tilde{R}_O^{\tau \tau^\prime}$ and thus the source of the CLFV physics.

\begin{figure*}
    \centering
    \includegraphics[scale=0.48]{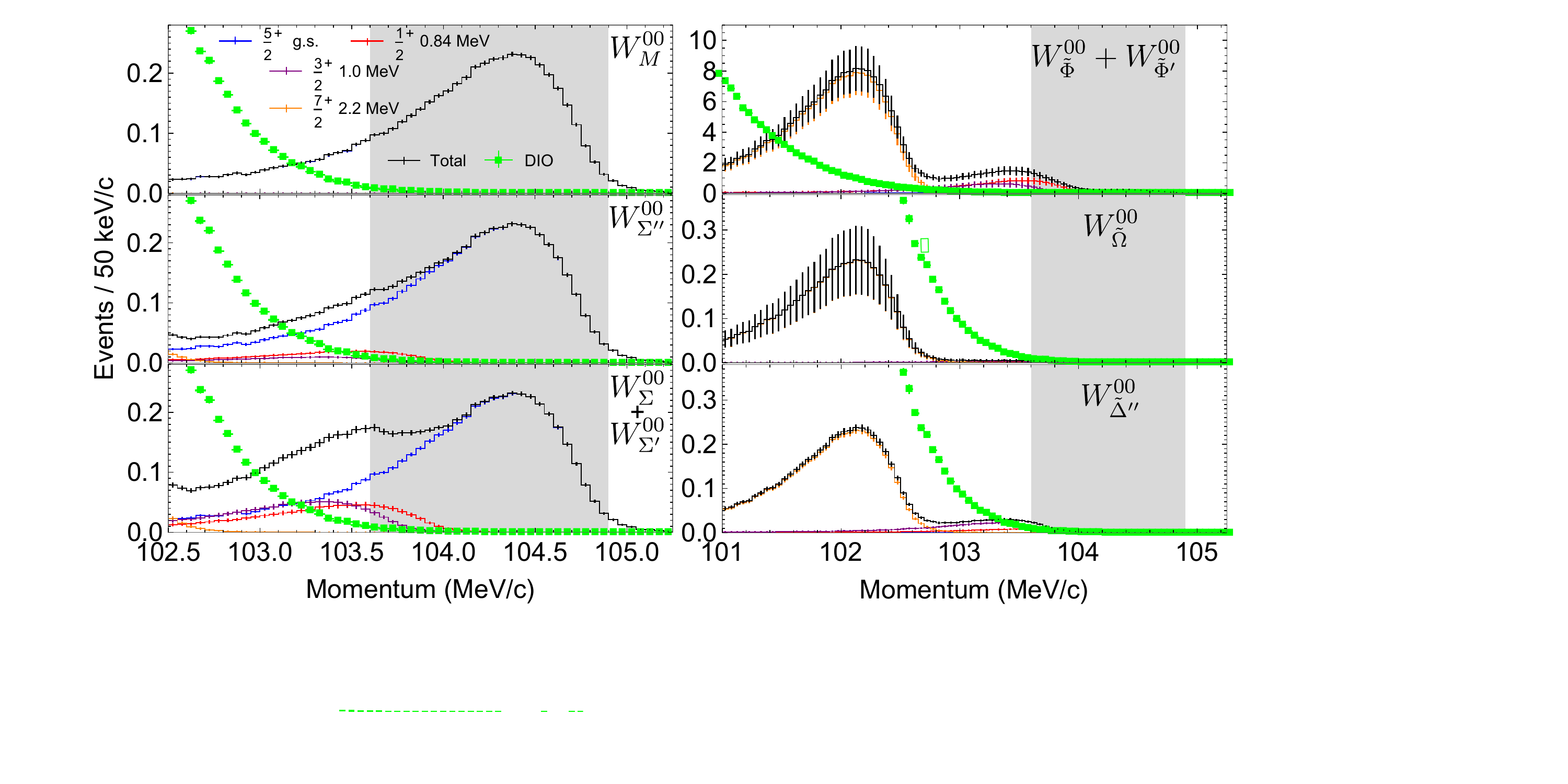}
    \caption{Expected electron counts in reconstructed momentum bins of width $50$ keV/c in Mu2e Run I for six NRET CLFV scenarios, mediated by charge ($M$), longitudinal spin ($\Sigma'')$, transverse spin ($\Sigma+\Sigma'$), transverse spin-velocity ($\tilde{\Phi}+\tilde{\Phi}'$), axial charge ($\tilde{\Omega}$), and longitudinal convection $(\tilde{\Delta}'')$. In each case, isoscalar coupling has been assumed.  The total CE signal (black) is separated into contributions from the nuclear ground state (blue) and first three excited states at 0.84 MeV (red), 1.0 MeV (purple), and 2.2 MeV (orange). Green squares denote the DIO background, which dominates all other sources \cite{Mu2e:2022ggl}. The gray shading indicates the region $103.60<q_\mathrm{rec}<104.90$ MeV/c where the Mu2e sensitivity has been optimized. Error bars combine the statistical uncertainty from the Mu2e detector response
    with the estimated nuclear theory uncertainty (see text).}
    \label{fig:mu2e_spec}
\end{figure*}


 Our near-endpoint electron spectrum calculations use results from recent Mu2e Monte Carlo simulations \cite{Mu2e:2022ggl}. We have added to the predicted Mu2e elastic CE signal the corresponding inelastic contribution, shifting the reconstructed electron momentum by the nuclear excitation energy $\Delta E_\mathrm{nuc}$ and rescaling the excited-state contribution by the
 branching ratio of Eq. \eqref{eq:rel_gamma_ratio}.  We take into account small phase-space differences and any CLFV effects that do not cancel
 in taking the ratio \cite{Haxton:2024amf}.  
 
 The needed $^{27}$Al response functions were evaluated for each of three selected $sd$-shell effective interactions \cite{PhysRevC.74.034315,bw} by performing full $sd$-shell diagonalizations. We employed harmonic oscillator Slater determinants with $b=1.84$ fm, an oscillator parameter chosen to reproduce the measured charge radius of $^{27}$Al . The standard deviations derived from these results were used as $1\sigma$ theory uncertainties, which were then folded in quadrature with CE spectrum statistical uncertainties. We stress that the nuclear uncertainty determined in this way is a minimum value, accounting for differences in the effective interactions, but not for common assumptions inherent in using the shell model.
 
 The shell-model effective interactions that we employ are tuned to reproduce the measured binding energies and excitation energies of $2s1d$-shell nuclei. The quality of the resulting predictions of beta-decay and electromagnetic transition moments throughout the shell has been thoroughly documented \cite{bw,BROWN1985347,PhysRevC.78.064302}. Specific
 tests of the wave functions for $^{27}$Al/$^{27}$Si are detailed in \cite{Haxton:2024amf}: The
 agreement between shell-model predictions and experiment is very good. 

\begin{table}[]
    \centering
    {\renewcommand{\arraystretch}{1.3}
    \begin{tabular}{l|c|c|c}
    \hline
    \hline 
     & \multicolumn{3}{c}{$R_{\mu e}(gs\rightarrow f)/R_{\mu e}(gs\rightarrow gs)$}\\
     Response & $f=1/2^+$ & $3/2^+$ & $7/2^+$\\
     \hline
    $W_M^{00}$ & $2.40(3)\times 10^{-4}$ & $4.4(2)\times 10^{-4}$ & $9.4(2)\times 10^{-4}$ \\
    $W_M^{11}$ & $8.2(5)\times 10^{-3}$ & $0.0113(5)$ & $0.019(1)$ \\
    $W_{\Sigma''}^{00}$     &  0.084(4) & 0.042(3) & 0.185(7)\\
    $W_{\Sigma''}^{11}$     &  0.081(4) & 0.055(4) & 0.194(10)\\
    $W_{\Sigma}^{00} + W_{\Sigma'}^{00}$ & 0.20(2) & 0.22(2) & 0.30(3) \\
    $W_{\Sigma}^{11} + W_{\Sigma'}^{11}$ & 0.22(3) & 0.22(3) & 0.33(4) \\
    $W_{\Phi''}^{00}$   &  $7(1)\times 10^{-4}$ & $8(1)\times 10^{-4}$ & $2.8(4)\times 10^{-3}$\\
    $W_{\Phi''}^{11}$   &  $6(3)\times 10^{-3}$ & $0.015(2)$ & $0.048(6)$\\
    $W_{\tilde{\Phi}}^{00} + W_{\tilde{\Phi}'}^{00}$ & $3.6(8)$ & $2.7(7)$ & $34(6)$\\
    $W_{\tilde{\Phi}}^{11}+ W_{\tilde{\Phi}'}^{11}$ & $7(3)\times 10^{-3}$ & $0.037(4)$ & $0.163(4)$\\
    $W_{\Delta}^{00} + W_{\Delta'}^{00}$ & $2.41(4)\times 10^{-3}$ & $2.6(2)\times 10^{-3}$ & $7.7(2)\times 10^{-3}$\\
    $W_{\Delta}^{11} + W_{\Delta'}^{11}$ & $2.1(2)\times 10^{-3}$ & $0.084(8)$ & $0.010(2)$\\
    \hline
    $W_{\tilde{\Delta}''}^{00}$ & $0.0361(8)$ & $0.097(4)$ & $1$\\
    $W_{\tilde{\Delta}''}^{11}$ & $0.062(6)$ & $0.12(1)$ & $1$\\
    $W_{\tilde{\Omega}}^{00}$  & $7(3)\times 10^{-3}$ & $0.016(6)$ & $1$\\
    $W_{\tilde{\Omega}}^{11}$ & $0.010(3)$ & $0.04(2)$ & $1$\\
    \hline
    \hline
    \end{tabular}}
    \caption{Relative $\mu\rightarrow e$ conversion strengths for transitions to the first three excited states of $^{27}$Al normalized to either the ground state (upper) or $7/2^+$ excited state (lower) (see text). Reported errors correspond to theory uncertainties estimated from the three nuclear shell-model calculations.}
    \label{tab:inelastic_responses}
\end{table}

If the elastic process is not forbidden, we normalize the CE spectrum by fixing the ground-state branching ratio to the fiducial value $R_{\mu e}(gs\rightarrow gs)=10^{-15}$. In cases where the elastic process is forbidden ($\tilde{\Delta}''$ or $\tilde{\Omega}$), we set the largest individual branching ratio (which, in all cases considered, is to the $7/2^+$ state at 2.2 MeV) to the fiducial value. This is the sensitivity anticipated for Mu2e Run I ($6\times 10^{16}$ stopped muons). Table \ref{tab:inelastic_responses} reports the relative strengths of the inelastic transitions for each nuclear response. The expected electron spectra for selected CLFV scenarios are shown in Fig. \ref{fig:mu2e_spec} and discussed below. Additional cases are explored in \cite{Haxton:2024amf}.

\textit{Coherent conversion.}
Many previous studies \cite{Kitano:2002mt,Cirigliano:2009bz,Crivellin:2017rmk,Bartolotta:2017mff,DAVIDSON2019380,HEECK2022115833,Cirigliano:2022ekw,Borrel:2024ylg} have focused on coherent $\mu\rightarrow e$ conversion, which can arise, for example, from a scalar or vector coupling of the leptons to quarks or a dipole coupling to the photon. We take the CLFV coupling to be isoscalar, to maximize the coherent enhancement of $W_M^{00}$. As no such coherence arises for inelastic transitions, the elastic contribution dominates the rate. From Table \ref{tab:inelastic_responses}, we see that transitions to excited states constitute $\approx 0.1\%$ of the total response, consistent with the expected elastic enhancement of  $\approx 0.43 A^2\approx 300$ (including effects of the elastic form factor). The top left panel of Fig. \ref{fig:mu2e_spec} shows that inclusion of excited-state contributions has no discernible effect on the simulated elastic spectrum of Ref. \cite{Mu2e:2022ggl}.

If the underlying CLFV charge coupling is isovector, then the coherence is lost: the scattering takes place on the unpaired nucleon. The leading multipole operator for the elastic process is monopole, $M_0$, while that for inelastic conversion is quadrupole, $M_2$. 
The resulting $q$-dependent dimensional suppression of inelastic responses is $\approx$ a factor of seven. The computed inelastic contribution is smaller, $\approx 3\%$ of the total, reflecting the specific 
nuclear structure of $^{27}$Al. Consequently, if experiment finds even a modest
inelastic contribution, the CLFV could not be entirely attributed to a charge interaction,
regardless of the charge's isospin couplings.

\textit{Spin-dependent conversion.} Spin-dependent operators, which couple primarily to the unpaired proton in $^{27}$Al, have been studied in the form $\vec{\sigma}_L \cdot \vec{\sigma}_N$ \cite{Cirigliano:2017azj,Davidson:2017nrp}.  In the NRET formalism two spin responses arise without velocity suppression, and three more arise when $\vec{v}_N$ is included to first order.  We discuss the former here.

Longitudinal coupling to spin, associated with the abnormal-parity operator $\Sigma_J^{\prime \prime}$, arises for pseudoscalar or axion-like-particle (ALP) CLFV exchanges \cite{Fuyuto:2024skf,Haxton:2024lyc}.  The contribution of the first excited state is roughly 9\% of the elastic response.  As the peak of the inelastic electron spectrum is displaced from the elastic peak, this produces a 40\% enhancement in the spectrum for $q_\mathrm{rec}\approx 103.5$ MeV/c, distinguishing this case from the charge responses discussed above. (See the left middle panel of Fig. \ref{fig:mu2e_spec}.) The leading operator for this transition is $\Sigma_3^{\prime \prime}$,
underscoring the importance of significant momentum transfer.

A transverse coupling to spin generates the electric and magnetic operators $\Sigma_J^\prime$ and $\Sigma_J$,  with parities $(-1)^{J+1}$ and $(-1)^J$, respectively.  The inelastic contributions are quite substantial, with each of the first three excited states contributing with strengths $\approx 20$--$35\%$ that of the ground state.
This creates a distinctive second peak in the CE spectrum near $103.5$ MeV/c, shown in the bottom left panel of Fig. \ref{fig:mu2e_spec} for isoscalar coupling.

\textit{Exotic responses.} 
Lastly, we consider several responses that depend explicitly on nuclear compositeness through their dependence on the inter-nucleon velocity operator $\vec{v}_N$ \cite{Rule:2021oxe,Haxton:2022piv}. When matched to a Lorentz-invariant, quark-level effective theory, some fine tuning is required to make $\vec{v}_N$-dependent operators leading \cite{Haxton:2024lyc}. From a bottom-up perspective, however, the associated response functions contain new CLFV information accessible to experiment. An interesting case is $W_{\tilde{\Phi}}^{00}+W_{\tilde{\Phi}'}^{00}$, generated from the transverse projection of $\vec{v}_N\times \vec{\sigma}_N$, a current that arises when the CLFV is mediated by tensor exchanges \cite{Haxton:2022piv,Haxton:2024lyc}. The transition to the $7/2^+$ state at 2.2 MeV is $\approx 40$ times stronger than the suppressed elastic transition, dominating the response, as shown in the top right panel of Fig. \ref{fig:mu2e_spec} (isoscalar coupling). The expected number of counts is high because of our adopted normalization to the ground-state rate.

The last two responses considered here are associated with CLFV operators that
can only be probed in inelastic transitions, as the elastic response vanishes.
 The response $\tilde{\Omega}$ is generated from interactions that couple to the nuclear axial charge, $\vec{v}_N \cdot \vec{\sigma}_N.$  Our calculations
 predict that $\gtrsim 95\%$ of the transition strength goes to the $7/2^+$ state at $2.2$ MeV. This places the signal --- shown in the right middle panel of Fig. \ref{fig:mu2e_spec} --- in a region where the DIO background is substantial, so that a background subtraction would be needed.  While the DIO shape is known well, ultimately the success of the subtraction will depend on statistical details of future experiments.


$\tilde{\Delta}''$ is the longitudinal projection of the nuclear convective current $\hat{q}\cdot\vec{v}_N$, which we assume is constrained by current conservation. 
Consequently, it can be eliminated in favor of the charge multipole
\begin{equation}
\tilde{\Delta}''_{JM}(q)=\frac{m_Nq_0}{q^2}M_{JM}(q),
\label{eq:siegert}
\end{equation}
where $q_0=-\Delta E_\mathrm{nuc}$ is the time component of the four-momentum transfer. (This rewriting exposes the explicit dependence on $\Delta E_\mathrm{nuc}$.  It also
reduces the impact of two-body corrections to operators, as these are second order in relativity for the charge operator.)
The response is dominated by the transition to the $7/2^+$ state, although the first two excited states also make modest contributions. This is shown in the bottom right panel of Fig. \ref{fig:mu2e_spec}, where the total electron spectrum is double peaked, with the primary response obscured by the DIO background.

\textit{Discussion.} We have demonstrated that the near-endpoint CE spectrum from elastic and inelastic $\mu \rightarrow e$ conversion in $^{27}$Al can vary significantly, depending on the underlying CLFV mechanism.  New information is provided by this spectrum, over and above that provided by the elastic rate alone.  While we have employed in our analysis Mu2e simulation data, our conclusions should apply equally  well to COMET. Our analysis is a sensitivity study, exploring separately several charge, spin, and convection current responses generated in NRET, assuming either isoscalar or isovector couplings. Clearly, multiple responses with arbitrary isospin couplings can contribute to total rates, requiring a more general NRET analysis.  Still, the basic conclusions reached here should hold up.  In particular, even a modest inelastic signal would rule out the most frequently explored model, CLFV generated entirely by a coherent charge coupling.

While the predicted counts in each 50 keV/c bin are typically small, our estimates are based on Mu2e Run I where $6\times 10^{16}$ muons will be captured. Over the total experimental lifetime, Mu2e is expected to stop $10^{18}$ muons in its $^{27}$Al target \cite{Bernstein_2019}, improving the statistics by more than an order of magnitude. Mu2e-II \cite{Mu2e:2018osu,Mu2e-II:2022blh}, a proposed extension leveraging proton beamline upgrades at Fermilab, could yield a further order-of-magnitude improvement. There is also the possibility that experiments can enhance their sensitivity to inelastic conversion by detecting the coincident low-energy photons emitted when the nucleus de-excites.\\

\begin{acknowledgments}
We are grateful to Pavel Murat, Kaori Fuyuto, and Liam Fitzpatrick for helpful discussions and to the Mu2e collaboration for making their simulation data available to us. WH acknowledges support by the US Department of Energy under grants DE-SC0004658, DE-SC0023663, and DE-AC02-05CH11231, the National Science Foundation under cooperative agreement 2020275, and the Heising-Simons Foundation under award 00F1C7. ER is supported by the National Science Foundation under cooperative agreement 2020275 and by the U.S. Department of Energy through the Los Alamos National Laboratory. Los Alamos National Laboratory is operated by Triad National Security, LLC, for the National Nuclear Security Administration of U.S. Department of Energy (Contract No. 89233218CNA000001).
\end{acknowledgments}

\bibliographystyle{unsrtnat}
\bibliography{mu2e_inelastic}

\end{document}